# Combined servo error pre-compensation and feedrate optimization – with application to a 3D printer and a precision motion stage

Heejin Kim[1] and Chinedum E. Okwudire[2]

*Abstract*–Servo error pre-compensation and feedrate optimization are often performed independently to improve the accuracy and speed of manufacturing machines. However, this independent approach leads to unnecessary trade-offs between productivity and quality in manufacturing. This paper proposes a novel linear programming approach for combined servo error pre-compensation and feedrate optimization, subject to contour error (tolerance) and kinematic constraints. The incorporation of servo error pre-compensation into feedrate optimization allows for faster motions without violating tolerance constraints. Experiments carried out on a 3D printer and precision motion stage are respectively used to demonstrate up to 43% and 47% reduction in cycle time without compromising part quality using the proposed compared to the independent approach.

## I. Introduction

A wide range of manufacturing machines use feed drives powered by computer numerical control (CNC) to generate motion commands. Two critical requirements of manufacturing are productivity and quality, which often involve a trade-off between speed and accuracy of feed drives [1]. This trade-off is typically handled, in practice, by maximizing speed so long as a an accuracy (tolerance) level is not violated. Servo errors are a major source of inaccuracy in feed drives. They can be caused by commanded motion (i.e., motion-induced servo errors) or disturbance forces like friction and manufacturing process forces. Motion-induced servo errors are very important in determining the trade-off between speed and accuracy because servo controllers always have limited bandwidth. This means that faster motion commands lead to larger servo errors. One way of reducing motion-induced servo errors is through pre-compensation (i.e., feedforward compensation). Knowledge of the machine's servo dynamics is used to modify the motion commands offline or online in the CNC interpolator to reduce servo errors. Examples of servo error pre-compensation (SEP) include zero phase error tracking controller [2], iterative method [3], path-modification via inverse dynamics [4], input shaper [5], analytical prediction and compensation of contour error [6], model predictive control framework [7], trajectory pre-filter [8], cross-coupled pre-compensation [9], mirror compensation with Taylor's expansion [10], adaptive cross-coupled prediction compensation [11], cross-coupled dynamic friction control [12], and filtered B splines [13,14]. However, available SEP approaches focus on reducing or minimizing servo error without trying to maximize feedrate subject to tolerance constraints.

On the other hand, there are numerous works on feedrate optimization (FO) subject to tolerance constraints. Traditionally, tolerance constraints are introduced implicitly into FO by imposing velocity, acceleration and jerk limits [15-18]. Some works have explicitly added tracking or chordal accuracy constraints to FO [19-25]. In practice, SEP and FO are combined by first performing FO and then applying the optimized results to SEP for error minimization, as shown by the one-way arrow between FO and SEP in Fig. 1(a). However, this independent (or sequential) approach could lead to sub optimality, as FO does not benefit from the reduction of error provided by SEP in maximizing feedrate. A better approach is to incorporate information from SEP into FO, as shown by the two-way arrow between FO and SEP in Fig. 1(b). To our best knowledge, such a combined technique for FO and SEP has not been explored in the open literature.

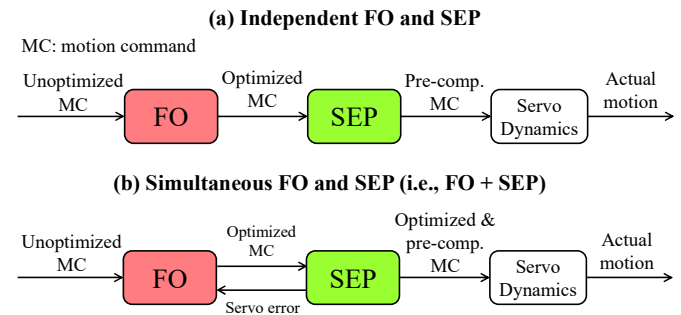

Fig 1: Comparison of (a). independent FO and SEP – standard practice – and (b). proposed concept of simultaneous FO and SEP (i.e., FO+SEP)

This paper proposes, for the first time, a linear programming (LP) approach for simultaneous FO and SEP. The systematic incorporation of SEP into FO expands the feasible region for

[1,2] All authors are with the Smart and Sustainable Automation Research Laboratory, University of Michigan, Ann Arbor, MI, 48109, USA {heejink,okwudire}@umich.edu.

FO, thus allowing for faster motions without violating tolerance constraints. In addition, the use of LP makes the proposed method computationally efficient and mathematically elegant. The outline of the paper is as follows: Section II contrasts time-based LP and available path-based LP [15,16] approaches for FO. It shows that time-based LP is similar to path-based LP in terms of computational efficiency but is superior to path-based LP in handling jerk constraints and incorporating servo dynamics into FO. Section III presents the proposed approach for simultaneous FO and SEP (i.e., FO+SEP) using time-based LP. Section IV validates the effectiveness of the proposed FO+SEP approach relative to FO in simulations and experiments carried out on a 3D printer and a planar motion stage. Conclusions and future work are presented in Section V.

## II. TIME-BASED VS PATH-BASED LP FOR FEEDRATE OPTIMIZATION

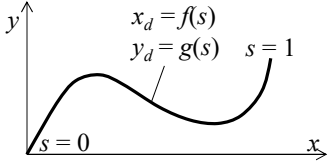

Fig 2: Parametric planar curve as function of path variable, $s$

Fig. 2 illustrates an arbitrary, curved path in the $x$-$y$ plane with path parameter $s \in [0,1]$. Note that $s$ is a function of time $t$ (i.e., $s = s(t)$). Let $x_d = f(s)$ and $y_d = g(s)$ denote a pair of parametric equations in $s$, representing the $x$ and $y$ components of desired position, respectively.

An increasingly popular approach is to perform FO via path-based LP, i.e., using $s$ as the independent variable [15,16]. In path-based LP, it is assumed that $s$ represents arc length, i.e., distance travelled along the curve, which is normalized by total travel length $L$. Let $x_d$ and $y_d$ be already known in the defined domain of $s$. Then, the kinematic limits $F_{max}$, $A_{max}$, and $J_{max}$ on feedrate, axis acceleration and axis jerk respectively, can be imposed as

$$L|\dot{s}| \leq F_{max}$$
$$\left|\frac{d^2 x_d(s)}{dt^2}\right| = |x_d''(s)\dot{s}^2 + x_d'(s)\ddot{s}| \leq A_{max} \quad (1)$$
$$\left|\frac{d^3 x_d(s)}{dt^3}\right| = |x_d'''(s)\dot{s}^3 + 3x_d''(s)\dot{s}\ddot{s} + x_d'(s)\dddot{s}| \leq J_{max}$$

for $\forall s$, where $x_d'(s)$, $x_d''(s)$, and $x_d'''(s)$ denote geometric derivatives of $x_d(s)$ with respect to $s$; $\dot{s}$, $\ddot{s}$, $\dddot{s}$ are tangential velocity, acceleration, and jerk, respectively. The $y$-axis acceleration and jerk limits are imposed in the same manner. Note that, instead of imposing feedrate limits as Eq. (1), axis velocity limits could be imposed in addition to axis acceleration and jerk limits [15,16].

To facilitate path-based LP, a new parameter $q = \dot{s}^2$ is introduced to remove the nonlinearity in Eq. (1). With $q$, the following substitutions hold:

$$\left. \begin{array}{l} \dot{s} = \sqrt{q}, \quad \dot{s}^2 = q, \quad \dot{s}^3 = q\sqrt{q} \\ \ddot{s} = \frac{1}{2}q', \quad \dddot{s} = \frac{1}{2}q''\sqrt{q}, \quad \dot{s}\ddot{s} = \frac{1}{2}q'\sqrt{q} \end{array} \right\} \quad (2)$$

With Eq. (2), the feedrate optimization with the same kinematic constraints is formulated as Eq. (3) for $\forall s$:

$$\begin{array}{l} \min_{q} -\int_0^1 q(s)ds \\ s.t. \; L\sqrt{q(s)} \leq F_{max} \\ \left| x_d''(s)q(s) + \frac{1}{2}x_d'(s)q'(s) \right| \leq A_{max} \\ \left| x_d'''(s)q(s) + \frac{3}{2}x_d''(s)q'(s) + \frac{1}{2}x_d'(s)q''(s) \right|\sqrt{q(s)} \\ \leq J_{max} \end{array} \quad (3)$$

Here, the feedrate constraint can be linearized by squaring both sides, and the $q'(s)$ and $q''(s)$ terms in acceleration can be linearized by B-spline parametrization of $q(s)$ with respect to $s$ [15,16]. However, because the $\sqrt{q(s)}$ term in the jerk limit is still nonlinear, $q(s)$ is replaced by a precomputed upper bound $q^*(s)$. One candidate for $q^*(s)$ is the solution obtained with only velocity and acceleration constraints [15,16] in Eq. (3). Then, the jerk constraint in Eq. (3) is reformulated using pseudo jerk $\bar{j}(s)$ as:

$$\left| x_d'''(s)q(s) + \frac{3}{2}x_d''(s)q'(s) + \frac{1}{2}x_d'(s)q''(s) \right|\sqrt{q^*(s)} = |\bar{j}(s)| \leq J_{max} \quad (4)$$

Although path-based LP is capable of imposing linear feedrate and axis acceleration constraints on FO, it cannot impose linear axis jerk constraints without the use of pseudo jerk, at the cost of optimality [15,16]. Moreover, because path-based LP uses $q$ as the independent variable, it is limited in its ability to accommodate servo dynamics, which uses time $t$ as the independent variable.

Therefore, in this work, we formulate a time-based LP approach for FO, adapted from the model predictive contour control framework proposed by Lam et al. [26]. Let $s(t)$ be discretized with fixed sampling interval, $T_s$, and expressed as a vector $\mathbf{s} = \{s(0), s(1), \ldots, s(N-1)\}^T$. Then, FO can be formulated as:

$$\min_{\mathbf{s}} \sum_{k=0}^{N-1} -s(k) \quad (5)$$
$$s.t. \; s(k-1) \leq s(k) \leq 1, \forall k = 1, 2, \ldots, N-1$$

The idea of Eq. (5) is that to minimize total time, the sum of $s(k)$ over $N$ time steps must be maximized – i.e., the path from $s(0) = 0$ to $s(N-1) = 1$ should be traversed as fast as possible, while satisfying the monotonicity and endpoint constraints on $\mathbf{s}$ in Eq. (5). In addition, it should satisfy kinematic constraints $F_{max}$, $A_{max}$, and $J_{max}$ on feedrate, axis acceleration and axis jerk, respectively:

$$L\frac{D[\boldsymbol{s}]}{T_s} \leq \boldsymbol{F}_{max}$$

$$\left|\frac{D^2[\hat{\boldsymbol{x}}_d]}{T_s^2}\right|, \left|\frac{D^2[\hat{\boldsymbol{y}}_d]}{T_s^2}\right| \leq \boldsymbol{A}_{max} \quad (6)$$

$$\left|\frac{D^3[\hat{\boldsymbol{x}}_d]}{T_s^3}\right|, \left|\frac{D^3[\hat{\boldsymbol{y}}_d]}{T_s^3}\right| \leq \boldsymbol{J}_{max}$$

Here, $D$ denotes finite difference operator, while $\boldsymbol{F}_{max}$, $\boldsymbol{A}_{max}$, and $\boldsymbol{J}_{max}$ are vectorized representations of the corresponding kinematic limits, and $\hat{\boldsymbol{x}}_d$ and $\hat{\boldsymbol{y}}_d$ are vectorized version of $\hat{x}_d(k)$ and $\hat{y}_d(k)$, respectively, similar to $\boldsymbol{s}$. This notation is maintained hereinafter. The terms $x_d = f(s)$ and $y_d = g(s)$ are generally nonlinear in $s$. Thus, at each time step $k$, they are linearized with linearization points $s^e(k)$ estimated from an initial unoptimized trajectory as

$$\hat{x}_d(k) = \left.\frac{\partial f(s)}{\partial s}\right|_{s=s^e(k)} \cdot (s(k) - s^e(k)) + f(s^e(k)) \quad (7)$$

and $\hat{y}_d(k)$ is obtained by linearizing $g(s)$ in the same manner.

To compare time-based and path-based LP, a circular toolpath with radius, $R = 5$ mm is employed, as illustrated in Fig. 3. The kinematic constraints are $F_{max} = 30$ mm/s, $A_{max} = 0.5$ m/s$^2$, and $J_{max} = 5$ m/s$^3$. For both path-based and time-based LP, $\boldsymbol{s}$ is discretized and represented using B-splines [16,20] as

$$\boldsymbol{s} = \boldsymbol{N}_s \boldsymbol{p}_s \quad (8)$$

where $\boldsymbol{p}_s$ is the control point vector of length $n_p$; $\boldsymbol{N}_s$ is the basis function matrix. By using $\boldsymbol{p}_s$ as the optimization variable in place of $\boldsymbol{s}$, the problem size is substantially reduced because $n_p \ll N$. Here, a 5$^{th}$ degree B-spline with uniform knot vector and $n_p = 40$ control points are used.

Time-based LP is initialized using an unoptimized trajectory generated using trapezoidal acceleration profile (TAP) [8] with the just-given kinematic limits. First, only the feedrate and acceleration limits are imposed on path-based and time-based LP. In this case, they yield almost the same feedrate profile, as shown in Fig. 4. As a result, their cycle time and computation time (using MATLAB® R2019a on a Windows PC with Intel Core i7-8750H CPU and 16 GB RAM) is similar, as summarized in Table 1. This shows that both methods have similar computational efficiency, under similar conditions. Next, the constraint on axis jerk and pseudo-jerk of $J_{max} = 5$ m/s$^3$ is introduced. Fig. 4 and Table 1 shows that the cycle time becomes 1.42 s for path-based LP and 1.25 s for time-based LP. This discrepancy shows the sub-optimality introduced by the pseudo-jerk relaxation. The computation time is also summarized in Table 1.

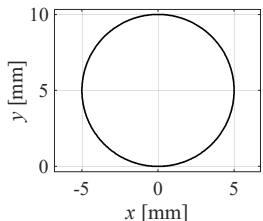

Fig 3: Desired toolpath

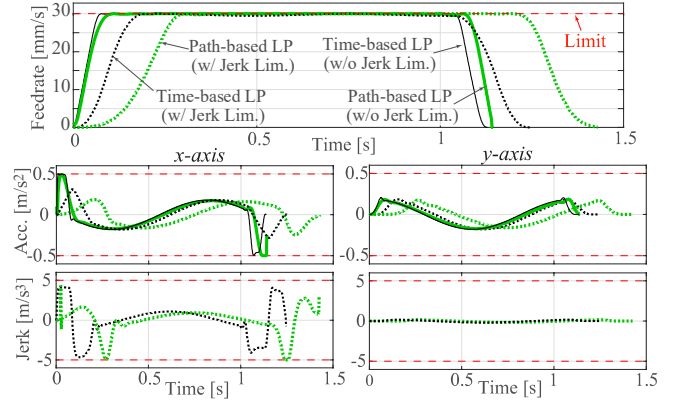

Fig 4: Feedrate, axis acceleration and axis jerk profiles of path- and time-based LP with and without jerk limits imposed

Table 1: Comparison of cycle and computation time of path- and time-based LP with and without jerk constraints imposed

| | FO algorithm | Cycle time [s] | Computation time [s] |
|---|---|---|---|
| w/o jerk constraints | Time-based LP | 1.13 | 0.75 |
| | Path-based LP | 1.14 | 0.77 |
| w/ jerk constraints | Time-based LP | 1.25 | 1.89 |
| | Path-based LP | 1.42 | 1.35 |

Time-based LP is also superior to path-based LP because it can incorporate any linear servo dynamics into FO. Conversely, path-based LP can only accommodate servo dynamics that are linear with regard to velocity and acceleration [19,20], without need for approximation. Given these advantages, the time-based LP formulated in this section is selected for the simultaneous SEP and FO approach proposed in the next section.

III. SIMULTANEOUS FO AND SEP USING TIME-BASED LP

*A. Framework of Simultaneous FO and SEP using Time-based LP*

Fig. 5 shows a block diagram of the proposed simultaneous SEP and FO. The idea is to impose contour error (tolerance) constraints on FO taking SEP into account. Contour error, denoted as CE hereinafter, has been selected as the accuracy index in FO because it directly impacts the ability of part quality to meet tolerance specifications in manufacturing [3-7,27]. However, because the proposed approach uses LP, CE must be estimated using linear dynamics. To do this, linearized desired $x$-axis position, $\hat{x}_d$, is used to generate modified position command $\hat{x}_{dm}$ using a SEP process represented by $C_x$. A linear model, $\hat{G}_x$, of the actual servo dynamics, $G_x$, is used to estimate the $x$-axis position as $\hat{x}$ and tracking error as $\hat{e}_x = \hat{x}_d - \hat{x}$. A similar process is followed to obtain $\hat{e}_y = \hat{y}_d - \hat{y}$, using $C_y$ and $\hat{G}_y$.

CE is defined as the orthogonal distance between an actual trajectory point at time $k$ and the reference toolpath [9], denoted as $\varepsilon_c(k)$ in Fig. 6. Approximate CE, denoted as $\hat{\varepsilon}_c(k)$, can be computed from the axis tracking errors using a linear estimation [27] as

$$\hat{\varepsilon}_c(k) = -\sin(\theta(k))\underbrace{(1 - \hat{G}_x C_x)\hat{x}_d(k)}_{=\hat{e}_x(k)} + \cos(\theta(k))\underbrace{(1 - \hat{G}_y C_y)\hat{y}_d(k)}_{=\hat{e}_y(k)} \quad (9)$$

where $\theta(k)$ is the angle of incline of the curve $(\hat{x}_d, \hat{y}_d)$ at time step $k$. The linear approximation of CE $\hat{\varepsilon}_c$ (including the effects of SEP using $C_x$ and $C_y$) is imposed as an additional constraint on the time-based LP formulation of Section II as:

$$|\hat{\boldsymbol{\varepsilon}}_c| = |-\sin(\boldsymbol{\theta})(\mathbf{I} - \hat{\boldsymbol{G}}_x \boldsymbol{C}_x)\hat{\boldsymbol{x}}_d + \cos(\boldsymbol{\theta})(\mathbf{I} - \hat{\boldsymbol{G}}_y \boldsymbol{C}_y)\hat{\boldsymbol{y}}_d| \leq \hat{\boldsymbol{E}}_{max} \quad (10)$$

where $\hat{\boldsymbol{E}}_{max}$ is the vectorized form of the maximum allowable approximate (i.e., linearized) CE, $\hat{E}_{max}$; $\boldsymbol{C}_x$, $\hat{\boldsymbol{G}}_x$, $\boldsymbol{C}_y$ and $\hat{\boldsymbol{G}}_y$ are matrix (lifted) versions of the corresponding system dynamics [28]; and $\mathbf{I}$ is the identity matrix. The implication is that a model of SEP is incorporated into FO, yielding FO+SEP. The optimized $x_d$ and $y_d$ from FO+SEP are then applied to the actual servo dynamics, $G_x$ and $G_y$, after being pre-compensated using $C_x$ and $C_y$, respectively. Note that if $C_x = C_y = 1$, then it means that no SEP is considered in FO.

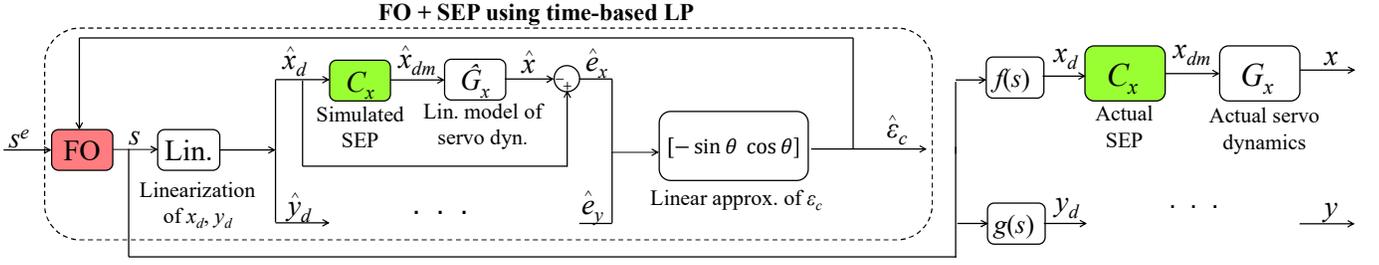

Fig 5: Block diagram of the proposed simultaneous SEP and FO method (y-component of SEP and servo dynamics omitted for simplicity)

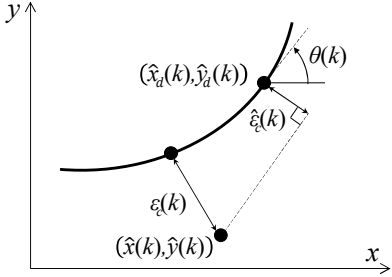

Fig 6: Contouring error $\varepsilon_c$ and its linear approximation $\hat{\varepsilon}_c$

### B. Realization of SEP in FO+SEP using filtered B splines

It is worth pointing out that $C_x$ and $C_y$ can be any linear SEP (feedforward tracking control) method, e.g., [2,4-6,8,9,11,13,14]. However, among the available linear SEP methods, the filtered B spline (FBS) approach [13,14] stands out because of its effectiveness and versatility in handling any type of linear system dynamics [14]. Therefore, it is selected for SEP in this paper.

The FBS approach parameterizes modified command $\boldsymbol{x}_{dm}$ (see Fig. 5) using B splines as $\boldsymbol{x}_{dm} = \boldsymbol{N}_x \boldsymbol{p}_x$, where $\boldsymbol{N}_x$ is the basis function matrix of degree $m$ and $\boldsymbol{p}_x$ is a vector of $n$ control points as

$$\underbrace{\begin{bmatrix} x_{dm}(0) \\ x_{dm}(1) \\ \vdots \\ x_{dm}(N-1) \end{bmatrix}}_{=\boldsymbol{x}_{dm}} = \underbrace{\begin{bmatrix} N_{0,m}(\xi_0) & N_{1,m}(\xi_0) & \cdots & N_{n-1,m}(\xi_0) \\ N_{0,m}(\xi_1) & N_{1,m}(\xi_1) & \cdots & N_{n-1,m}(\xi_1) \\ \vdots & \vdots & \ddots & \vdots \\ N_{0,m}(\xi_{N-1}) & N_{1,m}(\xi_{N-1}) & \cdots & N_{n-1,m}(\xi_{N-1}) \end{bmatrix}}_{=\boldsymbol{N}_x} \underbrace{\begin{bmatrix} p_x(0) \\ p_x(1) \\ \vdots \\ p_x(n-1) \end{bmatrix}}_{=\boldsymbol{p}_x} \quad (11)$$

where $\xi \in [0,1]$ is the spline parameter, representing normalized time; it is uniformly discretized into $\xi_0, \xi_1, \ldots, \xi_{N-1}$. Each basis function $N_{j,m}(\xi)$ is defined as

$$N_{j,m}(\xi) = \frac{\xi - \bar{g}_j}{\bar{g}_{j+m} - \bar{g}_j} N_{j,m-1}(\xi) + \frac{\bar{g}_{j+m+1} - \xi}{\bar{g}_{j+m+1} - \bar{g}_{j+1}} N_{j+1,m-1}(\xi)$$

$$N_{j,0}(\xi) = \begin{cases} 1 & \bar{g}_j \leq \xi \leq \bar{g}_{j+1} \\ 0 & \text{otherwise} \end{cases} \quad (12)$$

where $j = 0, 1, \ldots, n-1$, and $\bar{g} = [\bar{g}_0, \bar{g}_1, \ldots, \bar{g}_{n+m}]$ is a normalized uniformly-spaced knot vector defined in $[0,1]$. Accordingly, the system output is expressed as $\boldsymbol{x} \approx \hat{\boldsymbol{G}}_x \boldsymbol{x}_{dm} = \tilde{\boldsymbol{N}}_x \boldsymbol{p}_x$, where $\tilde{\boldsymbol{N}}_x = \hat{\boldsymbol{G}}_x \boldsymbol{N}_x$ (i.e., $\boldsymbol{N}_x$ filtered by $\hat{\boldsymbol{G}}_x$). The tracking error is modeled as Eq. (13):

$$\boldsymbol{e}_x \approx \boldsymbol{x}_d - \boldsymbol{x} = \boldsymbol{x}_d - \tilde{\boldsymbol{N}}_x \boldsymbol{p}_x \quad (13)$$

Then, the least-squares solution for minimizing $\boldsymbol{e}_x^T \boldsymbol{e}_x$ yields optimal coefficients $\boldsymbol{p}_x^*$ as:

$$\boldsymbol{p}_x^* = \left(\tilde{\boldsymbol{N}}_x^T \tilde{\boldsymbol{N}}_x\right)^{-1} \tilde{\boldsymbol{N}}_x \boldsymbol{x}_d = \tilde{\boldsymbol{N}}_x^\dagger \boldsymbol{x}_d \quad (14)$$

where † represents pseudoinverse. Therefore, $\boldsymbol{x}_{dm} = \boldsymbol{N}_x \boldsymbol{p}_x^* = \boldsymbol{N}_x \tilde{\boldsymbol{N}}_x^\dagger \boldsymbol{x}_d$, which leads to $\boldsymbol{C}_x = \boldsymbol{N}_x \tilde{\boldsymbol{N}}_x^\dagger$. The same process is applied to $\boldsymbol{y}_d$. Accordingly, $\boldsymbol{C}_x = \boldsymbol{N}_x \tilde{\boldsymbol{N}}_x^\dagger$ and $\boldsymbol{C}_y = \boldsymbol{N}_y \tilde{\boldsymbol{N}}_y^\dagger$ are substituted into Eq. (10) to realize FO+SEP using the FBS approach.

## IV. EXPERIMENTAL VALIDATION

For validation of the proposed FO+SEP approach, two experimental setups are used. The first set of experiments, described in Section IV-A, is carried out on a desktop 3D printer commonly used for rapid prototyping. The second set of experiments, described in Section IV-B, is carried out on a linear motor driven planar motion stage typically used in industry for precision positioning. Demonstration of the proposed method on two experimental setups helps to show its versatility.

### A. Desktop 3D printer

#### A.1. Experimental setup

A Lulzbot Taz 6 3D printer is used, as shown in Fig. 7. The optimization algorithms are implemented on dSPACE DS1202 real-time control board running at 1 kHz sampling rate, connected to DRV8825 stepper motor drivers for $x$, $y$, $z$, and $e$-(extruder) axes stepper motors. ADXL335 accelerometers are attached on the build plate and extruder to measure $x$, $y$-axes acceleration.

To execute FO and FO+SEP with error constraints, the $x$ and $y$ axis servo dynamics of the printer must be measured in the form of frequency response functions (FRFs) and modeled, via curve fitting, as $\hat{G}_x$ and $\hat{G}_y$. Fig. 8 shows the measured and modeled FRFs of $x$- and $y$-axes of the printer. The input of each FRF are swept sine acceleration commands to the stepper motor, and output is relative acceleration between the build plate and nozzle measured using the two ADXL335 accelerometers. The discrete-time transfer function representation of $\hat{G}_x$ and $\hat{G}_y$ are shown in Eq. (15), where the open-loop bandwidth is located around 30 Hz for both axes.

Moreover, in recovery of $x$, $y$ axes displacement from acceleration measurements, a Luenberger state observer [29] is used. Observer gains are chosen such that the dynamics of the observer error (i.e., difference between estimated position using the linear system model in Eq. (15) and observed position) obtains global asymptotic convergence with observer frequency $f = 10$ Hz.

$$\hat{G}_x = \frac{0.021z^5 - 0.061z^4 + 0.044z^3 + 0.033z^2 - 0.056z + 0.012}{z^6 - 5.627z^5 + 13.38z^4 - 17.2z^3 + 12.6z^2 - 4.994z + 0.836}$$

$$\hat{G}_y = \frac{0.018z^5 - 0.053z^4 + 0.038z^3 + 0.027z^2 - 0.048z + 0.017}{z^6 - 5.648z^5 + 13.48z^4 - 17.4z^3 + 12.8z^2 - 5.093z + 0.856}$$

(15)

#### A.2. Benchmarking to Determine Approximate CE Limit

Unoptimized position commands generated using trapezoidal acceleration profile (TAP) [8] are used for benchmarking to determine suitable approximate CE limit to traverse a circle of 5 mm radius. Two sets of kinematic limits are used. They are:

- <u>Conservative</u>: $F_{max} = 30$ mm/s, $A_{max} = 0.5$ m/s$^2$, $J_{max} = 5$ m/s$^3$;
- <u>Aggressive</u>: $F_{max} = 50$ mm/s, $A_{max} = 10$ m/s$^2$, $J_{max} = 5000$ m/s$^3$,

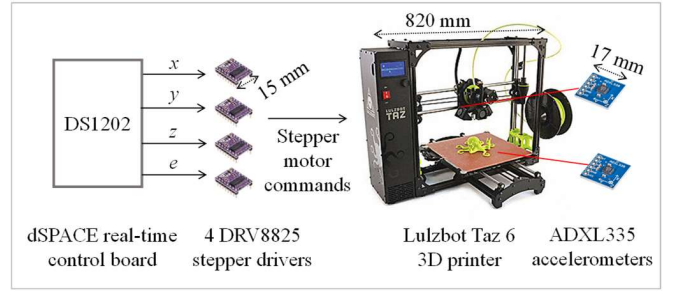

Fig 7: Experimental set up

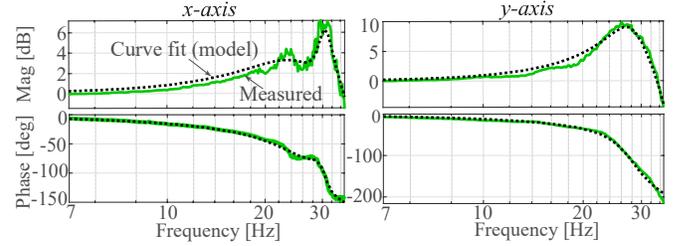

Fig 8: Measured and curve fitted FRFs of $x$ and $y$ axes of 3D printer

Fig. 9(a) shows the TAP feedrate profile generated using the conservative and aggressive kinematic limits; the acceleration and jerk profiles are omitted for the sake of brevity. Fig. 9(b) shows simulated (approximated) and actual (measured) CE profiles of the conservative and aggressive TAP position commands applied to the 3D printer. The simulations are performed using the curve fit linear dynamic model in Eq. (15). Conservative TAP yields maximum simulated and actual CEs of 14 μm and 54 μm, respectively. Conversely, Aggressive TAP yields maximum simulated and actual CEs of 30 μm and 141 μm, respectively. The reason for the discrepancy between the simulated and actual CEs is due to dynamics like friction and geometric errors not included in the linear model, as well as errors due to linear approximation of the CE in Eq. (9) and constraint equations in Eq. (6) and (10). As a result of these discrepancies between the linear dynamics/approximations and the actual dynamics, it is very important to determine the the approximate CE limits ($\hat{E}_{max}$), used in the proposed FO+SEP, that correspond to acceptable tolerance (i.e., actual CE). Prior work [30] has shown that the aggressive TAP results in poor print quality, while the conservative TAP (with maximum actual CE of 54 μm) yields acceptable print quality on the Taz 6 printer. Therefore, $\hat{E}_{max} = 14$ μm is selected as the approximate CE for LP-based optimization to help keep actual CEs close to the target 54 μm in reality.

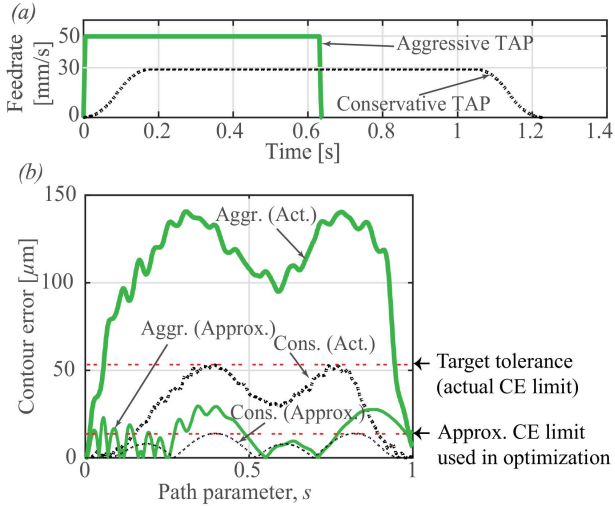

Fig 9: (a). Commanded feedrate and (b) simulated (approximated) and actual (measured) CE profiles of conservative and aggressive TAP motion commands

*A.3. Optimization Results using FO and FO+SEP*

We compare FO and the proposed FO+SEP with a goal to achieve similar accuracy as Conservative TAP in Fig. 9(b) with the shortest cycle time. To do this, the aggressive kinematic limits in Section IV.B are imposed on both FO and FO+SEP, together with an approximate CE limit of $\hat{E}_{max}$ = 14 μm using Eq. (10). For FO, $C_x = C_y = I$ (i.e., tolerance constraints are imposed without SEP). However, for FO+SEP, $C_x$ and $C_y$ are generated via the FBS approach described in Section III.B using a 5$^{th}$ degree B-spline with uniform knot vector and $n$ = 40 control points. Another 5$^{th}$ degree B-spline with uniform knot vector and $n_p$ = 40 control points is used to parametrize $s$ to reduce the problem size, as explained in Section II. Both the FO and FO+SEP cases are initialized using unoptimized TAP trajectories.

Fig. 10 shows the commanded feedrate, acceleration, and jerk profiles of FO and FO+SEP. Fig. 11 shows the simulated (approximated) and actual (measured) CE profiles. Both FO and FO+SEP enforce the approximate CE limit in the LP optimization, leading to the system staying close to the target in experiments. However, FO has to slow down because it hits the approximate CE limit while FO+SEP is able to stay very close to the maximum speed throughout the motion. As a result, FO+SEP completes the motion in 0.64 s, which is 43% faster than FO at 1.13 s, as summarized in Table 2. Note that implementing SEP after FO (i.e., independent approach) would not lower the cycle time of FO; it would only reduce the CE, which has little or no practical value if the desired tolerance has already been met. The computation time for FO+SEP is 1.8 s; FO's is much higher at 37.5 s because it is operating very close to the imposed error constraint.

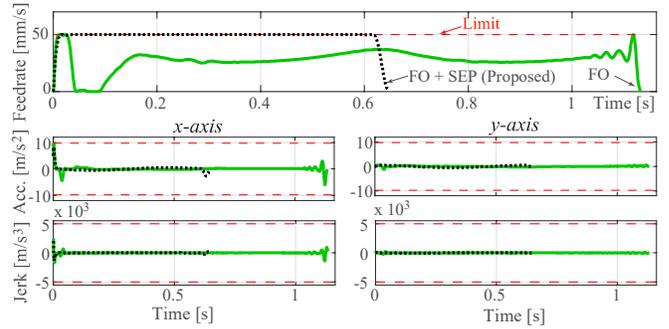

Fig 10: Feedrate, acceleration and jerk profiles of trajectories generated by FO and FO+SEP using aggressive kinematic limits and $\hat{E}_{max}$ = 14 μm

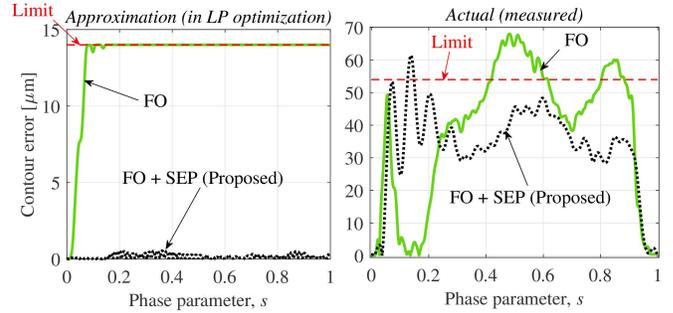

Fig 11: Simulated (approximated) and actual (measured) CE profiles of FO and FO+SEP

Table 2: Comparison of cycle and computation time of FO and FO + SEP

|  | Cycle time [s] | Computation time [s] |
| --- | --- | --- |
| FO | 1.13 | 37.5 |
| FO + SEP (Proposed) | 0.64 | 1.8 |

To further validate our findings, a cylinder of height 8.3 mm consisting of three concentric circular toolpath of radii 4.39 mm, 4.69 mm and 5 mm are printed using the same 3D printer, as shown in Fig. 12. Conservative and Aggressive TAP as well as FO and FO+SEP, as discussed above, are applied to each circular toolpath at each layer of the print. Fig. 13 shows the side and top view of the printed cylinders for the four cases. FO and FO+SEP save 10.9% and 50.5% in cycle time, respectively, compared to Conservative TAP, while maintaining similar surface quality. However, Aggressive TAP results in poor surface quality, though it takes a similar length of time as FO+SEP to print.

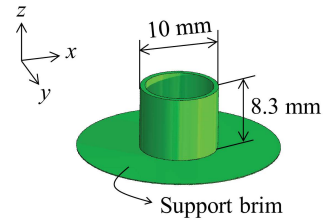

Fig 12: CAD model of cylinder

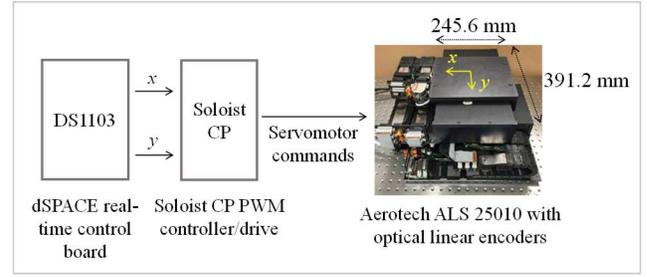

Fig 13: Side and top view of printed cylinders using Conservative TAP, Aggressive TAP, FO and FO+SEP on a support brim

### B. Precision Motion Stage

The performance of FO+SEP is also validated without the use of state observer, by testing on a precision motion stage with direct encoder feedback. This is to show that FO+SEP is applicable to various types of servo systems that suffer from limited bandwidth.

#### B.1. Experimental Setup

A biaxial linear-motor-driven motion stage (Aerotech ALS 25010) is used, as shown in Fig. 14. The optimization algorithms of FO and FO+SEP are implemented on dSPACE DS1103 real-time control board running at 1 kHz sampling rate, connected to Soloist CP controller/drive. Each axis is controlled by pre-tuned closed-loop P-PI controller and velocity feedforward. The planar motion stage is equipped with optical linear encoders with resolution 0.1 μm to provide position feedback on each axis.

As with the 3D printer, the servo dynamics of the $x$ and $y$ axis are measured in the form of FRFs and fitted as transfer functions. Fig. 15 shows the measured and modeled FRFs of each axis of the planar motion stage. The input of each FRF are position commands constructed by swept sine acceleration to the servomotor, and output is the position measured by encoders on each axis.

#### B.2. Benchmarking to Determine Approximate CE Limit

Unoptimized position commands generated using Conservative TAP [8] are used for benchmarking to determine suitable approximate CE error in traversing a circle of 5 mm radius. Fig. 16(a) shows the TAP feedrate profile generated using conservative kinematic limits of $F_{max}$ = 40 mm/s, $A_{max}$ = 0.4 m/s$^2$, $J_{max}$ = 4 m/s$^3$; the acceleration and jerk profiles are omitted for the sake of brevity.

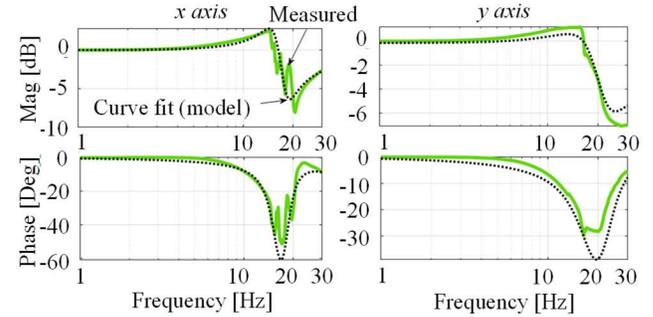

Fig 14: Experimental setup

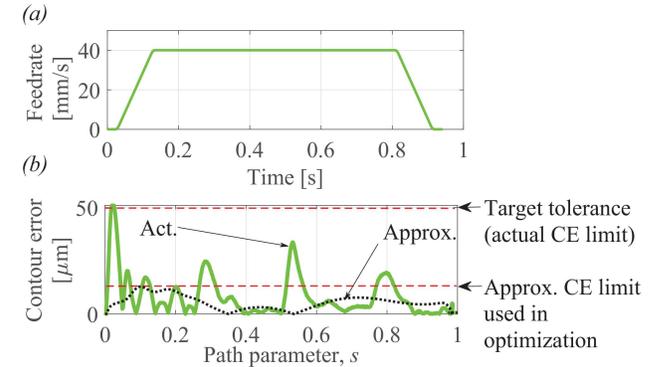

Fig 15: Measured and curve fitted FRFs of $x$ and $y$ axes of planar motion stage

Fig 16: (a) Commanded feedrate and (b) simulated (approximated) and actual (measured) CE profile of TAP motion commands

Fig. 16(b) shows simulated (approximated) and actual (measured) CE profiles of the TAP position commands applied to the planar motion stage. The simulations are performed using the curve fit linear dynamic model of Fig. 15. The conservative TAP yields maximum simulated and actual CE of 13 μm and 50 μm, respectively. Because the actual contouring accuracy of the TAP is considered to be satisfactory, $\hat{E}_{max}$ = 13 μm is selected as the approximate CE limit for LP-based optimization, to help keep CE close to the target 50 μm in reality.

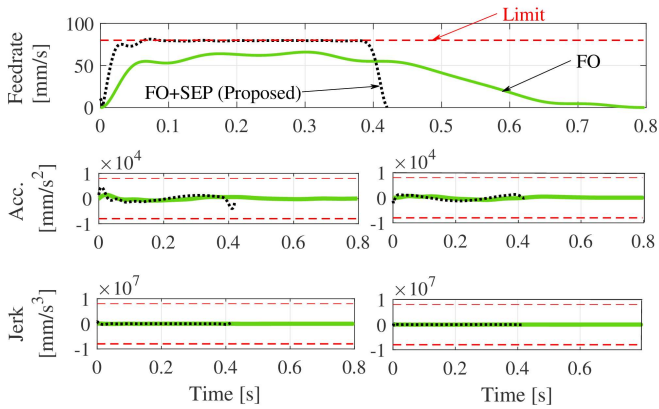

Fig 17: Feedrate, acceleration and jerk profiles of trajectories generated by FO and FO+SEP using aggressive kinematic limits and $\hat{E}_{max}$ = 13 μm

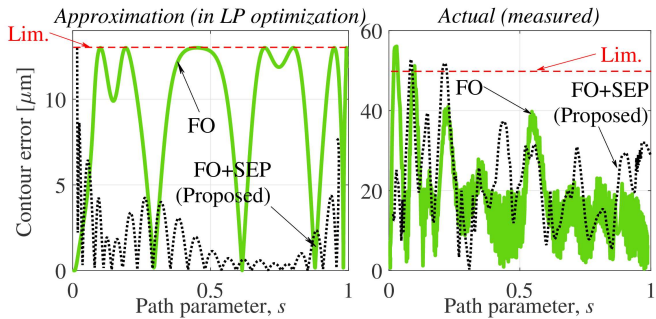

Fig 18: Simulated (approximated) and actual (measured) CE profiles of FO and FO+SEP

Table 3: Comparison of cycle and computation time of FO and FO + SEP

|  | Cycle time [s] | Computation time [s] |
|---|---|---|
| FO | 0.79 | 0.12 |
| FO + SEP (Proposed) | 0.42 | 0.04 |

### B.3. Optimization Results using FO and FO+SEP

We compare FO and the proposed FO+SEP with a goal to achieve similar accuracy as Conservative TAP in Fig. 16(b) with the shortest cycle time. To do this, aggressive kinematic limits are imposed on both FO and FO+SEP as: $F_{max}$ = 80 mm/s, $A_{max}$ = 8 m/s$^2$, $J_{max}$ = 8,000 m/s$^3$. In addition, approximate CE limit of $\hat{E}_{max}$ = 13 μm is imposed. In FO, $C_x = C_y = I$, and in FO+SEP, $C_x$ and $C_y$ are generated via the FBS approach described in Section III-B using a 5$^{th}$ degree B-spline with uniform knot vector and $n$ = 30 control points. To reduce the problem size, another 5$^{th}$ degree B-spline with uniform knot vector and $n_p$ = 30 control points is used to parametrize $s$. Both the FO and FO+SEP are initialized using unoptimized TAP trajectories.

Fig. 17 shows the commanded feedrate, acceleration, and jerk profiles of FO and FO+SEP. Fig. 18 shows the simulated (approximated) and actual (measured) CE profiles. Both FO and FO+SEP enforce the tolerance in simulations, which enforces the experimental error close to the target. However, as was in the experiment with the 3D printer, FO has to slow down because it hits the CE limit while FO+SEP is able to stay very close to the maximum speed throughout the motion. Consequently, FO+SEP completes the motion in 0.42 s, which is 47% faster than FO at 0.79 s, as summarized in Table 3. Note that the cycle time of FO is only 0.15 s (i.e., 16.0%) faster than the conservative TAP. The computation time for FO+SEP is 0.04 s; FO's is higher at 0.12 s because it is operating very close to the approximate error limit imposed.

### V. CONCLUSION AND FUTURE WORK

This paper has introduced a new concept of simultaneous FO and SEP (i.e., FO+SEP), and proposed a novel approach for realizing FO+SEP using time-based LP.

A time-based LP approach, which uses time as the independent variable, is formulated and compared with commonly-used path-based LP. Time-based LP is preferable to path-based LP in two aspects: (1) axis jerk constraints can be imposed without the use of pseudo-jerk approximation, and (2) any general linear dynamics constraints can be incorporated. It is shown in the simulations that time-based LP provides an elegant and computationally efficient approach for FO+SEP.

Compared to the standard practice of performing FO and SEP independently, FO+SEP relaxes the error tolerance constraints imposed on FO, allowing shorter cycle times without violating tolerance constraints. Experiments carried out on a 3D printer and precision motion stage yielded up to 43% and 47% reduction, respectively, in cycle time using FO+SEP compared to FO, subject to the same tolerance and kinematic constraints.

Future work will explore limited-preview (windowing) [14,16] into FO+SEP to enable its application to longer trajectories. Implementation of sequential linear programming approach will also be studied to reduce the linearization error [31]. Lastly, incorporation of actuator (e.g., torque) limits [20] into FO+SEP will also be explored to help expand its performance benefits over FO.

### VI. ACKNOWLEDGEMENTS

This work is funded by the National Science Foundation's award #1825133: Boosting the Speed and Accuracy of Vibration-Prone Manufacturing Machines at Low Cost through Software.